\documentclass[pra, twocolumn, amsmath]{revtex4}

\begin{document} 

\title{Consistent measurement of a quantum dynamical variable using classical apparatus} 
\author{E. C. G. Sudarshan}
\affiliation{The University of Texas at Austin, Center for Particle Physics, 1 University Station C1602, Austin TX 78712} 
\email[email: ]{sudarshan@physics.utexas.edu}

\begin{abstract}
The claim that there is an inconsistency of quantum-classical dynamics \cite{terno04} is investigated. We point out that a consistent formulation of quantum and classical dynamics which can be used to describe quantum measurement processes is already available in the literature\cite{ecg76}. An example in which a quantum system is interacting with a classical system is worked out using this formulation.
\end{abstract}
\maketitle

Measurement of a quantum system has been a fundamental problem. Niels Bohr \cite{bohr28} and J. von Neumann \cite{neumann} had emphasized that measurements on a quantum system must ultimately have a classical apparatus which displays the values of the quantum variable. Despite many decades of study and a voluminous literature this requirement on the dynamical process of measurement has {\em not} received much attention; in fact the explicit solution to this problem more than quarter of a century ago \cite{ecg76,sherry78,sherry79,gautam79} has not been given the attention it deserved. Instead we find that instances of `re-inventing the wheel' are being repeated \cite{peres01,terno04}. 
 
The standard postulates of quantum mechanics state that only commuting quantum dynamical variables can be measured using an experimental setup. It would be inconsistent to `measure' non-commuting variable with {\em any} classical apparatus. Since we need to couple the two systems we must first view a classical system as a special kind of quantum system \cite{danieri62,danieri66,koopman31,ecg74} with additional variables that are not observable. The dynamical description must involve these additional variables. For {\em different} measurements the set ups and the coupling terms would be {\em different}.

For this purpose let us choose a very simple system as the classical one that performs the role of the apparatus. A free particle moving in one dimension with a finite mass is one such system with a natural Poisson bracket. The dynamical variables of the system are labeled $a$ and $b$ partly to emphasis that the description of the classical system need not necessarily be in terms of the usual momentum and position variables and that all that is really required of the dynamical variables is that there exists a Poisson bracket between them; 
\[ [b \,,\, a]_{_{PB}} =1,\]
and hence
\[  [f(a,b) \,,\, g (a,b)]_{_{PB}}= \frac{\partial f}{\partial b} \frac{\partial g}{\partial a} - \frac{\partial f}{\partial a} \frac{\partial g}{\partial b}.\]

Let us now view this as a system with two sets of variables $a$, $b$, $\tilde{a}$, $\tilde{b}$ where 
\begin{equation}
  \label{eq:1}
  \tilde{a} \equiv i[a \,,\, \cdot \;]_{_{PB}} \qquad , \qquad \tilde{b} \equiv i[ b \,,\, \cdot \;]_{_{PB}}
\end{equation}
are linear non-commuting variables with the following {\em commutation} relations between them \cite{danieri62,danieri66}
\begin{eqnarray*}
  \tilde{b}\tilde{a}-\tilde{a}\tilde{b} &=& i^2\Big([b\,,\,[a\,,\,\cdot \;]_{_{PB}}]_{_{PB}}-[a\,,\,[b\,,\,\cdot\;]_{_{PB}}]_{_{PB}} \Big) \\
&=&i \widetilde{[b \,,\, a]}_{_{PB}} ,
\end{eqnarray*}
where we have used the Jacobi identity.
\begin{eqnarray*} 
  \tilde{b}a-a\tilde{b} &=&i[b\,,\,a]_{_{PB}} ,\\
  \tilde{a}b-b\tilde{a}=&=& i[a\,,\,b]_{_{PB}}, \\
\tilde{a}a-a\tilde{a}=0 &, & \tilde{b}b-b\tilde{b} =0
\end{eqnarray*}
using 
\[ (\tilde{b}a-a\tilde{b})f=i[b\,,\,af]_{_{PB}}+ ia[b\,,\,f]_{_{PB}}. \]
The dynamical variables $a$ and $b$ of the system by themselves always commute, i.e.
\[ ab-ba=0. \]

The familiar equations of motion for the classical system can be obtained starting from the Hamiltonian, 
\[ h(b,a) = \frac{a^2}{2m}. \]
Then
\begin{equation}
  \label{eq:2}
  \dot{b} = \frac{\partial h}{\partial a} = \frac{a}{m} \qquad ; \qquad \dot{a} = - \frac{\partial h}{\partial b} = 0
\end{equation}
We can now choose to treat the classical system like a quantum mechanical system with non-commuting variables by defining a new Hamiltonian operator $\tilde{h}$ as
\[ \tilde{h}(a,b, \tilde{a}, \tilde{b}) = \frac{a \tilde{a}}{m} \]
The Heisenberg equations of motion corresponding to this Hamiltonian are
\begin{eqnarray}
  \label{eq:2a}
  \dot{a}&=&-i(a\tilde{h}-\tilde{h}a)=0 \nonumber \\
  \dot{b}&=& -i\frac{a}{m} \times (i[b\,,\,a]_{_{PB}}) =\frac{a}{m} \nonumber \\
  \dot{\tilde{a}}&=&0 \nonumber \\
  \dot{\tilde{b}}&=& \frac{a}{m} \widetilde{[b \,,\, a]}_{_{PB}}.
\end{eqnarray}

The variables $\tilde{a}$, $\tilde{b}$ are not observable but the equations of motion for the observable dynamical variables $a$ and $b$ are identical to the Hamilton's equations. We now proceed to see how this quantum mechanical description of the classical system can be used to look at the case where it is interacting with a quantum system.

As a quantum system we may choose a system with a complete set of dynamical variables, either a spin system or a harmonic oscillator. Let us choose the latter with Hamiltonian
\begin{equation}
  \label{eq:3}
  H(\hat{q},\hat{p}) = \frac{\hat{p}^2}{2\mu} + \frac{ \mu \omega^2 \hat{q}^2}{2} \quad , \quad \hat{q}\hat{p} - \hat{p}\hat{q} = i  
\end{equation}
and equally spaced energy eigenvalues $\frac{1}{2} \omega$, $\frac{3}{2} \omega$, $\frac{5}{2} \omega$, $\ldots$ Let us consider the measurement of the quantum momentum $\hat{p}$ by a particular coupling term
\[ g\hat{p} \tilde{a}. \]
The total Hamiltonian is then
\begin{equation}
  \label{eq:4}
  {\cal{H}}(\hat{q}, \, \hat{p}, \, b, \, a, \, \tilde{b}, \, \tilde{a} ) =   \frac{\hat{p}^2}{2\mu} + \frac{ \mu \omega^2 \hat{q}^2}{2} + \frac{a \tilde{a}}{m} + g \hat{p} \tilde{a}
\end{equation}
Then
\begin{eqnarray}
  \label{eq:5}
  \dot{\hat{q}} & = & -i(\hat{q}  {\cal{H}} - {\cal{H}} \hat{q} ) = \frac{\hat{p}}{\mu} + g \tilde{a} \nonumber \\
  \dot{\hat{p}} &=& \mu \omega^2 \hat{q} \nonumber \\
  \dot{b} &=& \frac{\tilde{a}}{m} + g \hat{p} \nonumber \\
  \dot{a} &=& 0.
\end{eqnarray}
Since this experiment is set up to {\em measure the momentum} $\hat{p}$ the equation of motion for $b$ has a term proportional to it, and this is an observable. It is true that $\dot{\hat{q}}$ contains unobservable quantities; but this is in accordance with Bohr's dictum that there is an {\em uncontrollable disturbance} in the conjugate variable $\hat{q}$ when $\hat{p}$ is observed. If we wanted to measure the energy $ \frac{\hat{p}^2}{2\mu} + \frac{ \mu \omega^2 \hat{q}^2}{2}$ we should have chosen a coupling term
\[ g \left(  \frac{\hat{p}^2}{2\mu} + \frac{ \mu \omega^2 \hat{q}^2}{2} \right) \tilde{b} . \]

The operator Hamiltonian ${\cal{H}}$ is {\em not} an observable \cite{ecg76} since it contains $\tilde{a}$, $\tilde{b}$. If we are set up to measure $\hat{p}$ we {\em cannot} measure $\hat{q}$ nor the oscillator energy. $H(\hat{q}, \, \hat{p})$. This is as it should be, since $\hat{p}$ does not commute with $H(\hat{q}, \, \hat{p})$.

In place of a harmonic oscillator we could consider any system, not necessarily described by canonical variables \cite{ecg76}. For a quantum spin system we would couple any {\em one} of the spin components coupled to the classical momentum $b$ by a coupling term $g \tilde{a} \hat{S}_3$. Then the equation of motion for $\hat{S}_3$ is free of unobservable quantities and the value of $\dot{b}$ contains $g \hat{S}_3$. A more systematic treatment of the Stern-Gerlach set up to measure any one component of the spin is available in the literature \cite{sherry79}.
 
In summary the measurement of a {\em commuting} set of quantum observables can be consistently described by a suitable coupling to a classical system. The variables which do not commute with the specified quantum observables nor the linear operators of taking classical Poisson brackets can be measured.

I am thankful to Anil Shaji for bringing the inconsistent consistency claims to my attention and for useful discussions.
\bibliography{meas}
\end{document}